\documentclass[12pt]{article}
\usepackage{amsmath}
\usepackage{graphicx,psfrag,epsf}
\usepackage{enumerate}
\usepackage{natbib}
\usepackage{url} 
\usepackage{hyperref}
\usepackage{booktabs,float}
\usepackage{xcolor}
\usepackage{xr}
\usepackage{setspace}
\newcommand{\blind}{1}

\begin{document}

\def\spacingset#1{\renewcommand{\baselinestretch}%
{#1}\small\normalsize} \spacingset{1}


\if1\blind
{
  \title{\bf De Novo Functional Protein Sequence Generation:
  Overcoming Data Scarcity through Regeneration and Large
  Models}
 	\author{Chenyu Ren$^a$,  Daihai He$^{a}$,
	and
	Jian Huang$^{a,b}$\thanks{Corresponding author. Email: j.huang@polyu.edu.hk}\\
	\vspace{0.2cm}
	{\footnotesize {{\it$^a$Department of Applied Mathematics, The Hong Kong Polytechnic University}}}\\
	{\footnotesize { {$\it^{b}$Department of Data Science and Artificial Intelligence, The Hong Kong Polytechnic University}}}
}
  \maketitle} \fi

\if0\blind
{
  \bigskip
  \bigskip
  \bigskip
  \begin{center}
    {\LARGE\bf De Novo Functional Protein Sequence Generation:
  Overcoming Data Scarcity through Regeneration and Large Models}
\end{center}
  \medskip
} \fi

\bigskip
\begin{abstract}
Proteins are essential components of all living organisms and play a critical role in cellular survival. They have a broad range of applications, from clinical treatments to material engineering. This versatility has spurred the development of protein design, with amino acid sequence design being a crucial step in the process. Recent advancements in deep generative models have shown promise for protein sequence design. However, the scarcity of functional protein sequence data for certain types can hinder the training of these models, which often require large datasets. To address this challenge, we propose a hierarchical model named ProteinRG that can generate functional protein sequences using relatively small datasets. ProteinRG begins by generating a representation of a protein sequence, leveraging existing large protein sequence models, before producing a functional protein sequence. We have tested our model on various functional protein sequences and evaluated the results from three perspectives: multiple sequence alignment, t-SNE distribution analysis, and 3D structure prediction. The findings indicate that our generated protein sequences maintain both similarity to the original sequences and consistency with the desired functions. Moreover, our model demonstrates superior performance compared to other generative models for protein sequence generation.
\end{abstract}

\noindent%
\vfill

\newpage
\spacingset{1.9} 
\section{Introduction}
\label{sec:intro}

Functional proteins are essential components of all living organisms and play a critical role in cellular survival. They have a broad range of applications, from clinical treatments, such as targeting cancer cells and facilitating gene therapy, to material engineering, where enzymes are used to degrade plastics.
Their significance has catalyzed advancements in de novo protein design, particularly in the creation of functional proteins. While traditional protein design methods have made significant progress over the past decades \citep{huang2016coming,gainza2016algorithms,pan2021recent}, they are not without their drawbacks. These methods often employ computational techniques like the Monte Carlo algorithm to deduce optimal sequences based on a given backbone scaffold, rather than specific functions. Such processes can be computationally intensive and time-consuming, especially when incorporating the functional attributes of proteins.

In contrast, deep learning has recently gained traction as a powerful and efficient approach to protein engineering. A notable example is AlphaFold \citep{jumper2021highly}, an AI model introduced in 2020, which has been hailed as a breakthrough in predicting protein three-dimensional structures, effectively addressing the protein folding problem. Nonetheless, the challenge of de novo protein design persists; it involves designing the primary structure, or amino acid sequence, of a functional protein. This sequence is the starting point for models like AlphaFold, as it dictates the protein's physicochemical properties, molecular function, and ultimately, its three-dimensional structure.

The rapid advancements in generative models, such as Generative Adversarial Networks (GAN) \citep{goodfellow2014generative}, Variational Autoencoders (VAE)  \citep{kingma2014auto},
 and diffusion models \citep{ho2020denoising, song2021scorebased},
 have revolutionized the fields of image and text generation. These models also hold significant promise for protein design, with various proposed models leveraging them to craft distinct functional amino acid sequences.  \cite{repecka2021expanding} introduced an unconditional GAN for generating the malate dehydrogenase (MDH) enzyme, and \cite{kucera2022conditional} employed a conditional GAN to create different functional proteins based on given Gene Ontology (GO) annotations.

However, a major challenge in the supervised training of these generative models is the scarcity of labeled data.  \cite{lyu2023proteinvae} highlighted the limited availability of unique hexon sequences, with only 711 unique full-length sequences from the UniprotKB database, which hampers efficient training. A similar issue arises in the design of lysozyme, an enzyme crucial for its antibacterial, anti-inflammatory, and antiviral properties in medicine, cosmetics, and food preservation \citep{yang2017shell,shaku2020peptidoglycan,chipman1969mechanism,osserman2012lysozyme}. Most natural lysozyme is derived from bird species, which poses a problem for individuals allergic to egg white \citep{vocadlo2001catalysis,fremont1997prevalence}. The UniprotKB database contains only about 2650 unique full-length sequences for Lysozyme (C for 1163, G for 1490), and some lack essential keywords. This limited dataset poses a significant challenge for effectively training generative models.

The recent advent of ChatGPT has demonstrated the potential of large language models (LLMs), particularly through the application of representation leraning and transfer learning. This approach involves pre-training on a vast corpus of unlabeled data followed by fine-tuning on a smaller set of labeled data, effectively addressing the challenge of data scarcity.
The application of LLMs to protein sequences has gained considerable attention in recent years. AlphaFold \citep{jumper2021highly}, developed by DeepMind, AlphaFold uses deep learning and attention mechanisms to predict protein structures with high accuracy. ProtBert \citep{brandes2022proteinbert} trained on large protein sequence databases to perform various prediction tasks, such as  prediction of  protein functions. ESM (Evolutionary Scale Modeling) \citep{lin2022language}  
leverages transformer architectures to capture evolutionary patterns in protein sequences.
\cite{madani2023large} developed a large language model capable of generating amino acid sequences based on specific keywords for various protein families. However, the reliance on keywords presents a challenge, as they are not always available or complete in the UniProtKB database. Additionally, re-training a pre-trained model to accommodate different protein descriptions can be prohibitively expensive.

To overcome these challenges, we propose a hierarchical model, ProteinRG,  inspired by regeneration learning \citep{tan2023regeneration}.
Our model can be trained on a limited dataset comprising various functional protein sequences.
We evaluate the generated results from three distinct perspectives:  multiple sequence alignment, t-SNE distribution analysis, and 3-D structures corresponding to the sequences.
Furthermore, we employ the maximum mean discrepancy (MMD) statistic to assess the distributional similarity and the mean reciprocal rank (MRR) to evaluate the conditional consistency of the generated protein sequences with the actual sequences. Our two-stage model demonstrates superior performance compared to other one-stage generative models for protein sequence design.

\section{Overview}
\label{sec:overview}
\begin{figure*}
    \centering
    \includegraphics[width=0.9\linewidth]{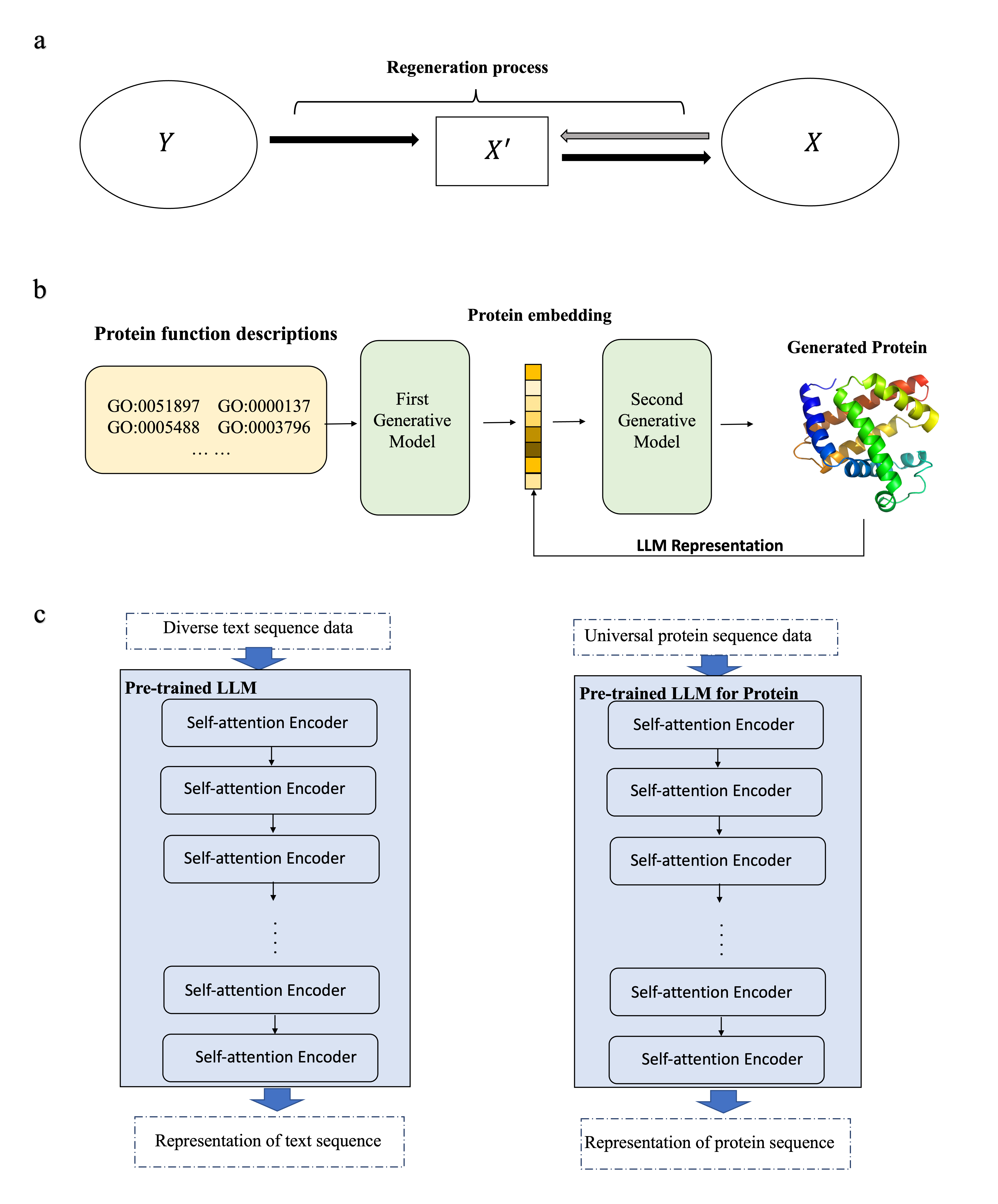}
    \caption{a. The regeneration process is a concept designed for mapping complex distributions from $Y$ to $X$. The process begins by transforming $X$ into an abstract representation $X'$, followed by regenerating $X'$ from $Y$, and ultimately generating $X$ from $X'$.
b. Our proposed ProteinRG for functional protein sequence generation adheres to this regeneration process. It commences with the generation of a representation of protein sequences, which is derived from a large protein sequence model, and subsequently generates the actual protein sequence.
c. On the left, we have the pre-trained large language model for natural language processing, while on the right is the pre-trained large model specifically for proteins. Both models utilize multiple self-attention encoder blocks from the transformer architecture.}
    \label{fig:rgl}
\end{figure*}

Functional protein sequence design can be conceptualized as a task of conditional data generation, where the objective is to learn a mapping that can be used to sample new data from the conditional
distribution of target data $X$ with condition $Y.$ The concept of regeneration learning, a recently proposed framework \citep{tan2023regeneration, ramesh2022}, demonstrates significant promise in learning complex distributions. In this context, for a given data point $X$, let $X^{\prime}$ represent an intermediate abstraction of $X$. When compared to $X$, $X^{\prime}$ is of a lower dimensionality, making it easier to map. Furthermore, $X^{\prime}$ retains more information pertinent to conditional generation than $Y$. The overall methodology of regeneration learning is depicted in Figure \ref{fig:rgl}a.

Inspired by the concept of regeneration learning,
we propose ProteinRG, a hierarchical model specifically designed for generating protein sequences, even when data availability is limited. Our model employs an intermediate representation of protein sequences, which acts as the product of the initial generation phase and the precursor for the subsequent phase. ProteinRG's efficacy stems from its use of pre-trained large-scale protein sequence models for learning a representation, which adapted techniques from the domain of LLMs.
This strategy effectively addresses the challenge of complex generative learning tasks in scenarios where data is scarce.

ProteinRG is a deep generative learning-based model that
efficiently generates functional protein sequences with minimal paired data.
The model is composed of three innovative modules that work in concert to capture latent features of protein sequences and regenerate them with desired functionalities (Fig. \ref{fig:enter-label}).

\begin{figure*}
    \centering
    \includegraphics[width=1\linewidth]{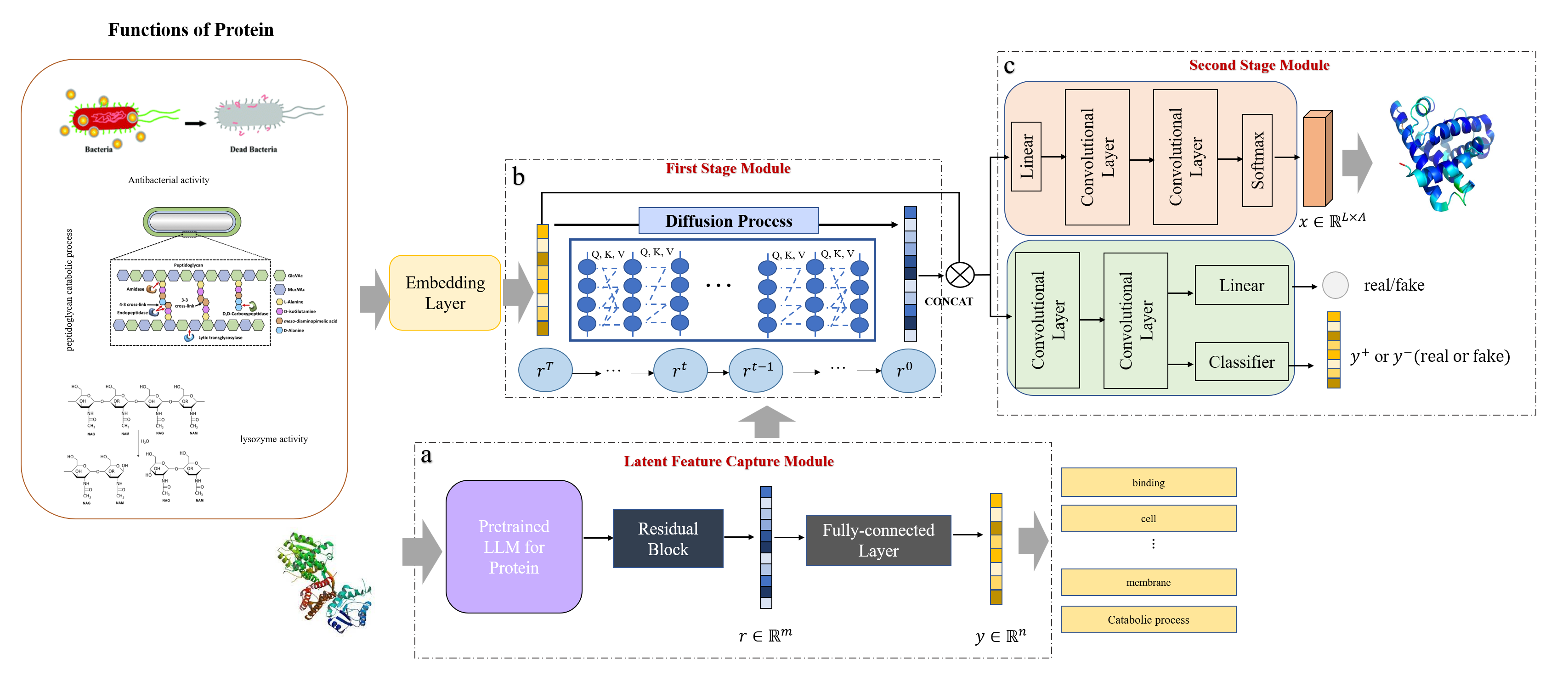}
    \caption{Overview of the proposed hierarchical functional protein generation model. Our model comprises three modules: (a) the latent feature capture module, (b) the first-stage generative module, and (c) the second-stage generative module. It begins by generating a latent representation of a protein sequence before producing a functional protein sequence.}
    \label{fig:enter-label}
\end{figure*}

Specifically, ProteinRG initially generates a latent representation of protein sequences, followed by the generation of functional protein sequences. We integrate a pre-trained large language model
for protein sequence data to learn the latent representation of proteins \citep{9477085, hsu2022learning,lin2022language, rives2019biological}, combined with a two-stage generative model enhanced by an auxiliary classifier network for protein sequence design.
A distinctive feature of ProteinRG is its capability to produce high-quality protein sequences while circumventing the need for extensive datasets. Such datasets are frequently scarce or non-existent for numerous protein types.

\section{Method}
Let $X$ denote an amino acid sequence and $Y$ represent the annotation label or sequence type corresponding to $X.$ The problem of functional protein sequence generation can be formulated as sampling from the conditional distribution of $X$ given $Y,$ denoted as $P(X|Y).$

Suppose the length of the amino acid sequence is $L$, the number of annotations is $A,$  and the dimension of $Y$ is $d.$
We use a one-hot vector of length $A$ to represent the amino acid type. Therefore, the dimension of $X$  is $L\times A.$
There are two main challenges in solving this problem. First, the conditional distribution is unknown, and only a random sample from the joint distribution of $(X,Y) $ is available. Second, the dimensionality of $X$ is high. For protein sequence data, the length of an amino acid sequence is typically $L=2042,$  and since there are 21 types of amino acids, $A=21.$ Thus, the dimension of $X$ is $2048\times 21=43,008.$
Training a generative model to sample from a high-dimensional distribution with limited data is extremely challenging. To address this issue, we propose a regeneration method that utilizes an existing large protein sequence model to mitigate the problem of data scarcity.

Let $R$ denote a representation of the protein sequence $X$, capturing its essential features in an $m$-dimensional numerical vector.
In this work, we take $R:=R(X)$ to be a deterministic function that maps $X$ from a high-dimensional space, specifically $(L \times A)$ dimensions, to a lower-dimensional, $m$-dimensional space. This function is designed to retain critical information while reducing dimensionality. Typically, $m$ is significantly smaller than $L \times A$. Given that $L \times A$ is often large, directly learning a distribution in a high-dimensional space is challenging with a limited amount of data. To address this challenge, we leverage a useful property of the conditional distribution that utilizes the representations $R$ and express the conditional distribution $P(X \mid Y)$ as follows:
\begin{equation}
\label{rg1}
    P(X \mid Y) = P(X, R(X) \mid Y) = P(X \mid R(X), Y) P(R(X) \mid Y),
\end{equation}
where the first equality is justified by the properties of conditional probability.
This relationship allows us to decompose the high-dimensional conditional distribution into more manageable components.

 Therefore, our proposed ProteinRG model aims to learn the conditional distribution $P(X\mid Y)$ in a hierarchical manner based on the equation (\ref{rg1}). As presented in Steps 1 and 2,
 rather than directly learning $P(X \mid Y)$, ProteinRG instead learns to generate from two simpler conditional distributions on the right side of equation (\ref{rg1}).  This hierarchical generative approach is analogous to the method employed in text-conditional image generation \cite{ramesh2022} and can be considered a form of regeneration learning (\cite{tan2023regeneration}).

An overview of ProteinRG is depicted in Figure \ref{fig:enter-label}. The model comprises three modules that computationally implement the proposed method. The statistical basis of the method is
expressed in equation (\ref{rg1}) .

\begin{itemize}
    \item \textit{Module 1: Protein sequence representation learning}: In this module, we
     fine-tune a pre-trained large protein sequence model to learn a representations $R$ of a high-dimensional amino acid sequence $X$: $R = R(X)$.

\item \textit{Module 2: Learning the conditional distribution $P(r\mid y)$}:
In this module, we train a generative model $G_1$ to learn the conditional distribution $P(R \mid Y).$ This module is used to generating latent representations for a given annotation label.
That is, for a given label $Y=y$, this module allows us to sample  $r \sim P(R\mid Y=y).$

\item \textit{Module 3: Learning the conditional distribution $P(X\mid R, Y)$}: In this module,
we train a second generative model $G_2$
for learning the conditional distribution $P(X \mid R, Y)$. This module is use to generate amino acid sequences conditional on $(R, Y)=(r, y)$: $x\sim P(X \mid R=r,Y= y),$ where $r$ is generated from the generative model in Module 2.
\end{itemize}

After we have trained the model as described above,  the generating process is then:
for a given annotation $Y=y$,
\begin{itemize}
\item Step 1:  generate $r \sim P(R\mid Y=y)$,
\item Step 2: generate $x \sim P(X|R=r, Y=y).$
\end{itemize}

A key feature of our approach is that we fine-tune a pretrained large protein sequence model for learning the representation function $R,$ which captures more information than the annotation labels about the structural details of amino acid sequences. This makes the sampling process from $P(R|Y)$ and then from $P(X|R,Y)$ more effective than directly sampling from $P(X|Y)$.

\section{Description of model details}
In this section, we present a detailed description of the three modules comprising ProteinRG.
Our training dataset consists of pairs $\{x,y\} = \{ (x_{i},y_{i}): i = 1,2,...,N \}$, where $x_i$ represents the $i$th amino acid sequence, $y_i$ corresponds to its annotated label, and $N$ is the number of sequences.

\subsection{Module 1: Protein sequence representation learning}
In the latent feature capture module, we learn an effective representation of a protein sequence using
a pre-trained large language model (LLM) specifically designed for protein sequence data.

Recent advancements have demonstrated that LLMs develop emergent capabilities as they scale, transcending mere pattern recognition to facilitate higher-level reasoning and the generation of realistic images, text, and various sequential data forms. These models epitomize the concept of representation learning, where a task-agnostic latent representation is learned through unsupervised or self-supervised training on extensive collections of unlabeled data. Subsequently, this representation can be fine-tuned with paired data for a variety of downstream tasks, showcasing the potential for application across diverse domains. Large language models (LLMs) based on transformer architecture, employ an attention mechanism to discern interaction patterns among each single item within the input text or images.

The attention mechanism is a pivotal concept in the field of neural networks, particularly in sequence-to-sequence models. It was introduced to enhance the performance and efficiency of models by allowing them to focus on specific parts of the input sequence when generating each part of the output sequence. The incorporation of attention mechanisms has significantly improved the performance of models in tasks such as machine translation, text summarization, and more, by enabling them to dynamically focus on relevant parts of the input. Traditional models, such as the encoder-decoder architecture, often struggle with long sequences due to the fixed-size context vector. The attention mechanism addresses this limitation by creating dynamic context vectors for each output element. This is achieved through the computation of attention weights, which determine the relevance of each input element to the current output element being processed. The detail of transformers and attention mechanism are explained in the Supplementary Materials.

In this work, we utilize ESM-2, a state-of-the-art large language model specifically designed for capturing representations of proteins. This model has been pre-trained with 15 billion parameters on a dataset comprising over 60 million protein sequences from the UniRef50 database, making it one of the largest protein language models evaluated to date.
We fine-tune the ESM-2 model using training data. The downstream task is to classify protein functions. The model receives a set of protein sequences as input and produces multi-class Gene Ontology (GO) annotations as output.

Specifically, we design a network architecture that includes a residual block and fully connected layers, tailored for the supervised learning task. We extract the latent representation of protein sequences from the residual block within the classification network. To fine-tune the pre-trained
large protein sequence model ESM-2, we use Cross Entropy (CE) as the loss function, which is defined as
\begin{equation}
   L_{\text{CE}}= -{E}_{(X,Y) \sim p(x,y)}\left[p\left(X, Y\right) \log p\left(Y \mid X\right)\right].
\end{equation}
The training process involves adjusting the parameters of the classification network as well as certain parameters of the pre-trained model. The remaining parameters in the ESM-2 model are kept fixed.

We  note that using the output from the pre-trained model directly, without passing it through the classification network, can also serve as a representation of the protein sequences. This is useful
in a semi-supervised setting, where the dataset includes protein sequences whose annotations are
not available.

\subsection{Module 2: Learning the conditional distribution $P(R\mid Y)$}

We use a denoising diffusion probabilistic model (DDPM, \cite{ho2020denoising})
and a transformer-based network \citep{attension2017}
to learn a generative model for the conditional distribution $P(R \mid Y).$
This mode is used to generate the latent representations of protein sequences.

DDPM is a parameterized Markov chain that is trained using variational inference to generate samples that match the target data distribution within a finite number of steps \citep{ho2020denoising}. Standard diffusion models incorporate two key phases: the forward process, which is used for training, and the reverse process, which is utilized for sampling. The critical aspect of training diffusion models is to estimate a denoising function for predicting noise that is added at each time step. The loss function for estimating the denosing function is expressed as a mean square error between the actual noise and the predicted noise. More details are given in the Supplementary Materials.

\begin{figure}[H]
	\centering
	  \includegraphics[width=0.9\linewidth]{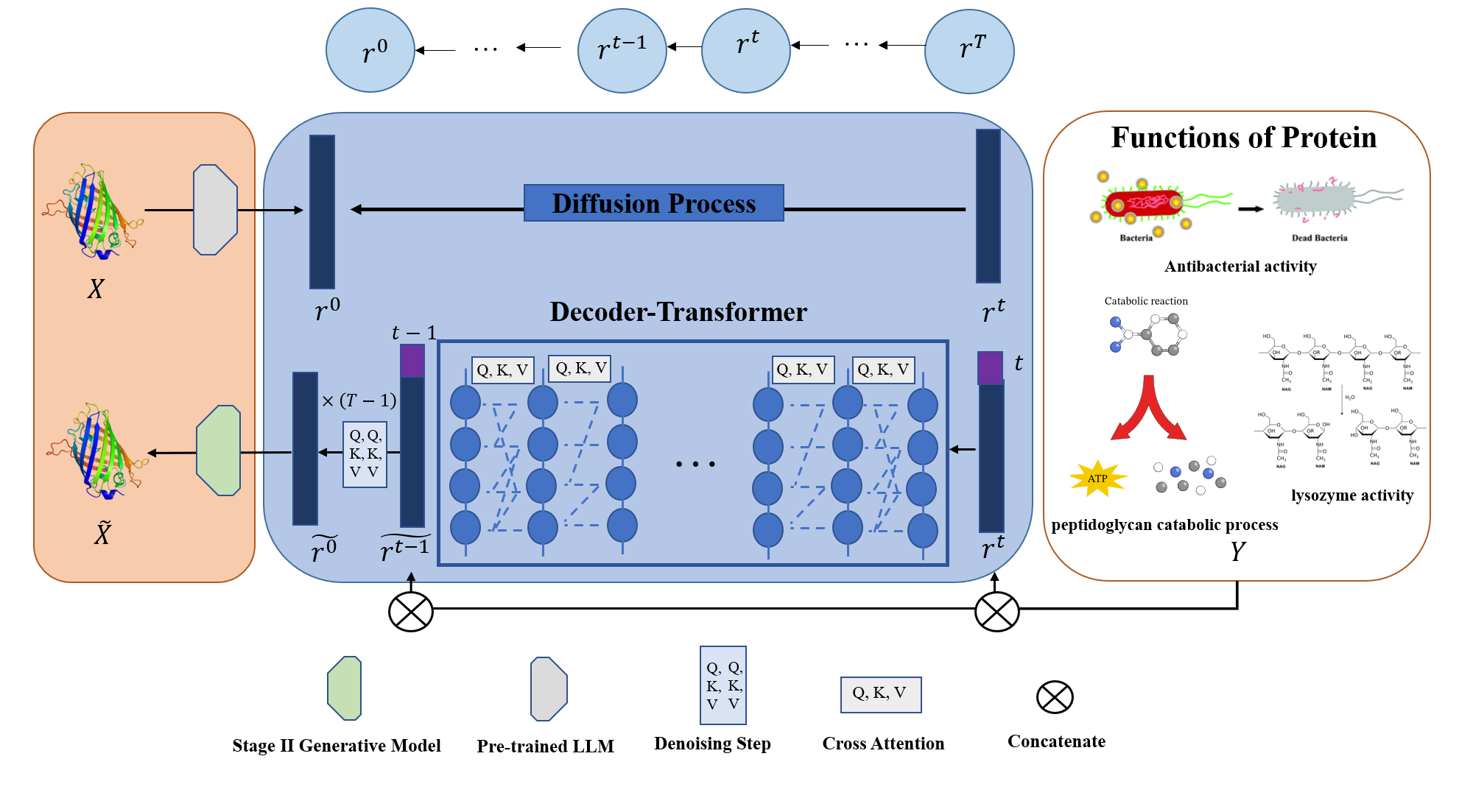}
        \caption{Overview of Module 2. Learning protein sequence representations given annotations.}
        \label{fig_E1}
\end{figure}

In our specific application of the diffusion model, we have found it more effective to predict the final latent representation $r$ directly, rather than predicting the noise in the forward process as in the original diffusion model. Therefore, we train the model using the following loss function,
\begin{equation*}
    L_{\text{diff}}(\theta) = {E}_{t \sim[1, T], R \sim q_t}\big[\big\|R_\theta (R, t, y)-r\big\|^2\big],
\end{equation*}
where $T$ represents the total number of diffusion time steps,  $t$ is encoded using a cosine embedding, and $y$ is a given condition. The function $R_\theta$ is a decoder-only transformer \citep{hsu2022learning} designed to generate the latent representation of a protein sequence, conditioned on the given annotation $y$.

To enhance the performance of sampling, we employ classifier-free guidance during training, which involves randomly omitting the condition $Y$ in 10\% of the time steps.
While the latent feature capture module constructs a link between the protein sequence $X$ and its latent representation $R$, the first stage generative module builds a mapping from functions of protein sequence $X$ to its latent representation $R$. We design a classifier-free guidance diffusion model (\cite{ho2020denoising}) to generate the representation $R$ conditioned on the functions $Y$.

\subsection{Module 3: Learning the conditional distribution $P(X\mid R, Y)$}

Building upon the latent embedding $r$ obtained from the latent feature capture module and the representation generated from the first stage generative module, we use a Conditional Wasserstein Generative Adversarial Network (CWGAN) with Gradient Penalty (CWGAN-GP) \citep{gulrajani2017improved}
in the generation of protein sequences. CWGANs are composed of two neural networks: a generator
and a discriminator. The training of CWGAN-GP involves a minimax game where the generator
seeks to minimize the Wasserstein distance with a gradient penalty between the real and generated data distributions, while the discriminator aims to maximize this distance.

 In the CWGAN-GP, both the generator and the discriminator receive a condition as part of their input. In this work, the condition is specified as the latent representation $R$ and annotation $Y$ of a protein sequence. To simplify the notation, let $U=(R, Y)$. The whole loss function can be written as  follows:

\begin{equation}
\label{wgan1}
V(G,D) =  \underset{(Z,U) \sim {P}_Z{P}_U}{E}[D(G (Z,U),U]-\underset{(X,U) \sim {P}_{(X,U)}}{E}[D({X,U})]
+\lambda \text{Pen}(D),
\end{equation}
where $Z \sim P_Z=N(\mathbf{0}, \mathbf{I})$ is a Gaussian noise vector,
$\lambda \ge 0$ is a tuning parameter,  and
$$\text{Pen}(D) = \underset{(\hat{{X}},{Y} )\sim {P}_{\hat{{X}}}{P}(Y)}{{E}}\left[\left(\left\|\nabla_{\hat{{X}}} D_{\eta}(\hat{{X}},Y)\right\|_2-1\right)^2\right]
$$
is a penalty term to ensure the Lipschitz regularity of the discriminator
function $D$.

To enhance conditional consistency and address the issue of low intra-class diversity, we propose the integration of an auxiliary discriminative classifier capable of handling multiple classes. Drawing inspiration from ADCGAN \citep{hou2022conditional}, which introduces an auxiliary discriminative classifier to distinguish between real and generated data using binary labels, we adapt this approach to our multi-class labeled data. We employ mutually exclusive one-hot encoding for multi-class generated data, ensuring clear discrimination of the generated data from the real data.

\begin{figure}
    \centering
    \includegraphics[width=0.90\linewidth]{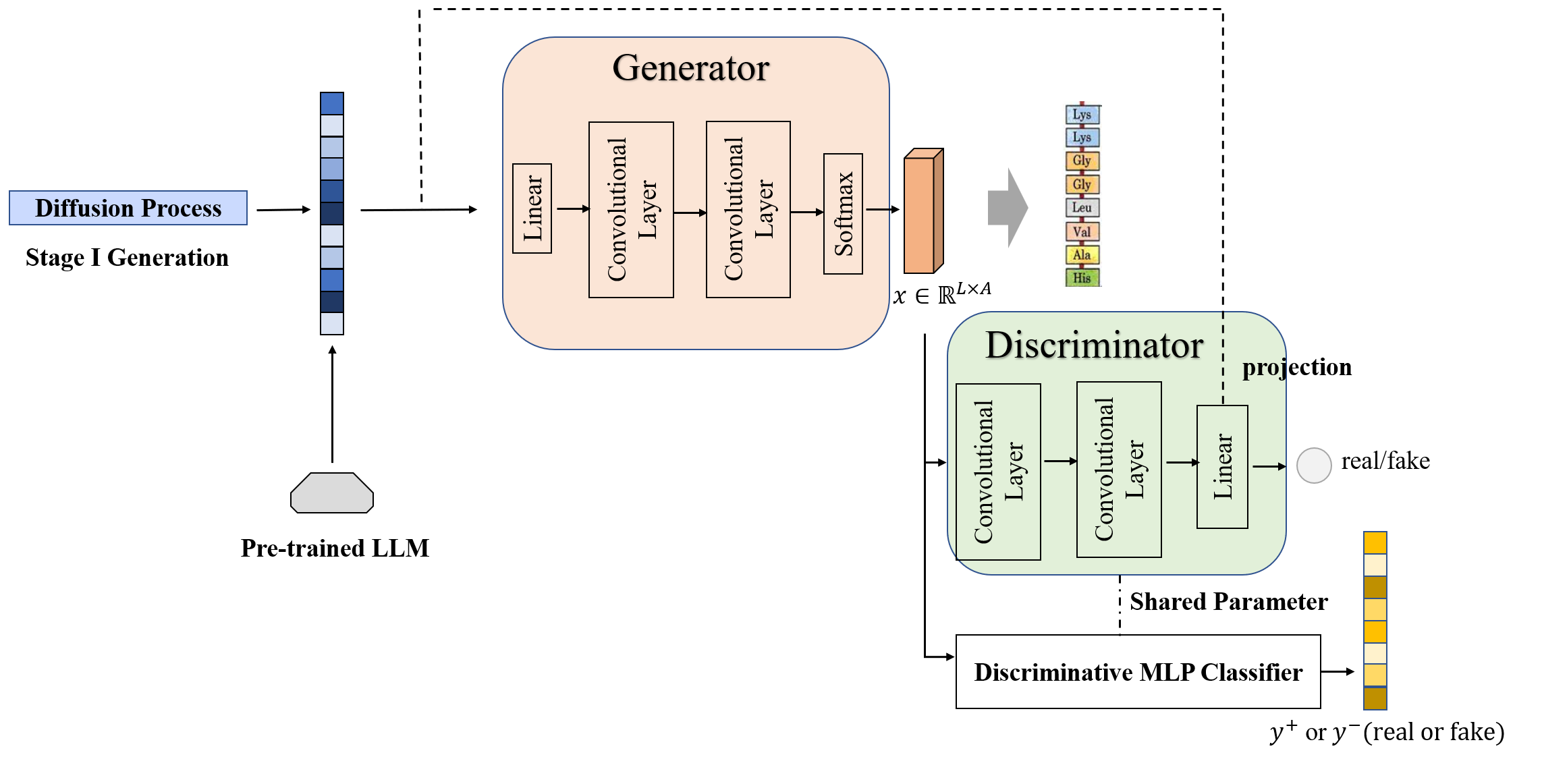}
    \caption{Overview of Module 3: a conditional Wasserstein Generative Adversarial Network with an auxiliary multi-class discriminative classifier is trained to generate protein sequences given the
    annotation.}
    \label{fig:enter-label}
\end{figure}

The objective functions for the generator, the discriminator, and the discriminative classifier are
defined as follows:
\begin{align*}
\max _{D, C_{\mathrm{d}}} \Big\{V(G, D)+\beta & \cdot \{{E}_{(X, Y) \sim P_{X, Y}}[\log C_{\mathrm{d}}(Y^{+} \mid X)]
+E_{(X,Y) \sim P_{X, Y}}[\log C_{\mathrm{d}}(Y^{-} \mid X)]\}\Big\},\\
 \min _G \Big\{V(G, D)-  \beta & \cdot
  \{{E}_{(X,Y) \sim P_{X, Y}}[\log C_{\mathrm{d}}(Y^{+} \mid X)]-{E}_{(X, Y) \sim P_{X, Y}}[\log C_{\mathrm{d}}(Y^{-} \mid X)]\}\Big\},
\end{align*}
where $V(G, D)$ represents the CWGAN-GP objective function defined
in (\ref{wgan1}),
$Y^+$ is a binary label indicating real protein sequence or generated one,
$Y^{-}$ is a one-hot vector representing the types of the generated multi-class protein sequences,
and a hyperparameter $\beta$ is introduced to balance the influence of the discriminator and the classifier.

\section{Training}
Our network training is conducted on four Nvidia GeForce A100 GPUs. The training process is divided into two  stages, with the first stage module and the second stage being trained independently. We utilize the pre-trained large language model ESM-2 (esm2\_t12\_35M\_UR50D), which is equipped with 12 self-attention layers and encompasses 35 million parameters. The output generated by this pre-trained large language model is a one-dimensional continuous vector. Throughout the fine-tuning process, we keep the ESM-2 weights fixed, with the exception of those in the final layers.

To enhance the training of the first stage module, we augment the representation to 512 dimensions by appending zeros. Consequently, the output from the latent feature capture module is transformed into a 512-dimensional vector. This adjustment ensures that the initial stage of the network is better equipped to process and learn from the input data, facilitating more effective training and potentially leading to improved performance in subsequent tasks.

In our experiments, we found that batch sizes of both 64 and 128 work best for the generation tasks. Also, among various learning rates, 1e-5 seems to be the best fit. The different values are summarized in the Table\ref{table-hyper} below.
\begin{table}
    \centering
    \setlength{\tabcolsep}{12mm}{
    \begin{tabular}{@{}cc@{}}\toprule
    Hyperparameter  &  Values\\ \midrule
    Learning Rate  & 0.001, 0.0004, 0.0001, 0.00004, 0.00001 \\
    Batch Size   & 32, 64, 128 \\
    Latent Dimension  & 10, 320, 640 \\
    $\beta$   & 1, 10, 100, 135, 175, 200\\
   \bottomrule
    \end{tabular} }
\caption{\centering Different Values of the hyperparameters for the Model}
\label{table-hyper}
\end{table}

\section{Applications}
\label{sec:results}

In this section, we apply the proposed method to two protein sequence datasets available from the UniProt Knowledgebase \citep{uniprot2019uniprot}: the Lysozyme sequence dataset and the Malate dehydrogenase (MDH) sequence dataset. These datasets consist of protein sequences and
protein keywords. They can be accessed online at \url{https://www.uniprot.org/}.

\subsection{Lysozyme sequences}
The lysozyme sequence dataset from the UniProtKB database
consists of  two most prevalent types: Lysozyme C and Lysozyme G. Within this database, we identified 1097 annotated sequences for Lysozyme C and 1668 for Lysozyme G. We excluded sequences containing non-standard amino acids (U, J, Z, O, B, X) to ensure data quality. Consequently, our training dataset consists of 2249 lysozyme sequences. Additionally, we set aside 516 lysozyme sequences for validation purposes, maintaining the same class distribution as observed in the original dataset.

\begin{figure}[H]
    \centering
    \includegraphics[width=0.9\linewidth]{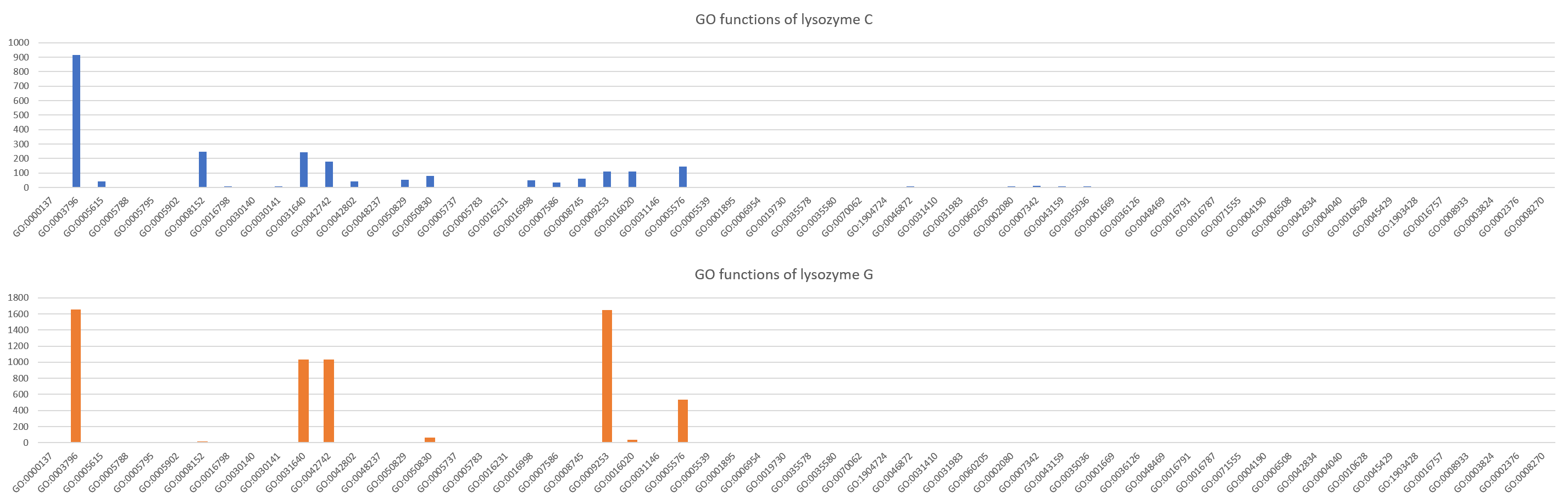}
    \caption{GO annotations of lysozyme C and lysozyme G. The upper panel is for lysozyme C and the lower one is for lysozyme G.}
    \label{fig:data}
\end{figure}

Due to the occasional absence of protein keywords in the UniProtKB database, we have utilized Gene Ontology (GO) annotations to ascertain the functions of lysozyme sequences. The Gene Ontology knowledge base is recognized globally as the definitive resource for the comprehensive description of gene functions. GO annotations provide insights into a gene's molecular function, its cellular location, and the biological processes or pathways it is involved in.

For our dataset, we identified 60 GO annotations related to lysozyme sequences, encompassing functions such as lysozyme activity (GO:0003796), defense response to bacterium (GO:0042742), killing of cells of another organism (GO:0031640), metabolic process (GO:0008152), digestion (GO:0007586), among others. Figure \ref{fig:data} presents the detailed GO functions for both types of lysozyme sequences. From this figure, it is evident that both Lysozyme C and Lysozyme G share the function of lysozyme activity (GO:0003796). However, distinct functions are also observed; for instance, Lysozyme G is predominantly associated with the peptidoglycan catabolic process (GO:0009253), which involves the breakdown of peptidoglycans.

We evaluate the generated lysozyme sequences from three distinct angles: at the one-dimensional level using multiple sequence alignment (MSA), at the two-dimensional level through t-SNE distribution analysis \citep{vanDerMaaten2008}, and at the three-dimensional level via AlphaFold2 \citep{jumper2021highly}. The MSA is performed using Clustal Omega \citep{sievers2011fast}, and we further compute Shannon's entropy for the MSA results at each position to assess sequence conservation.
For dimensionality reduction, we employ the t-SNE technique, utilizing the scikit-learn t-SNE package with its default settings. We initially embed the protein sequences using ESM-2, which yields a 480-dimensional representation. This high-dimensional data is then condensed to a two-dimensional space for visualization purposes.

 \begin{figure*}
    \centering
    \includegraphics[width=0.80\linewidth]{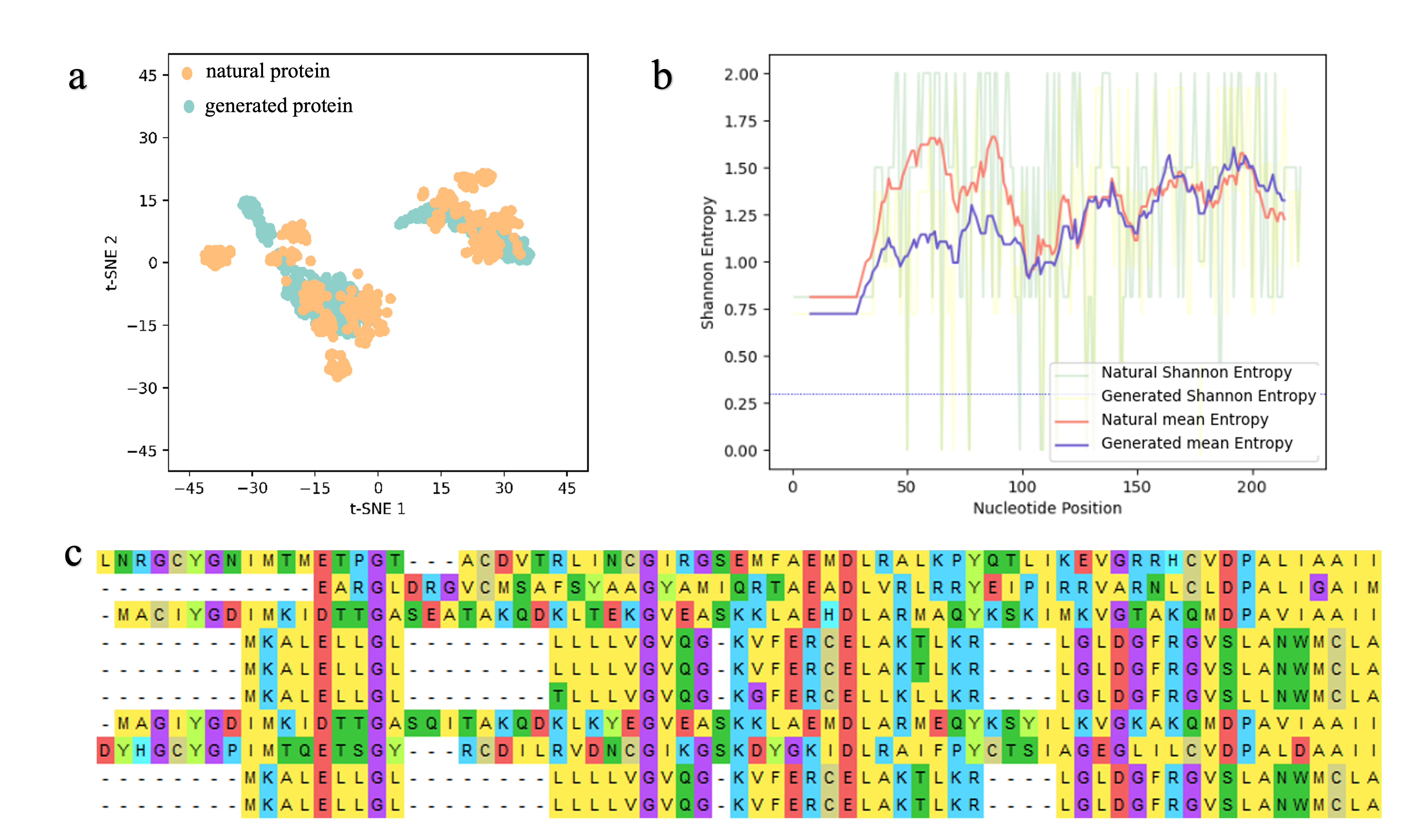}
    \caption{a.t-SNE visualization of natural protein from validation dataset and generated protein. The yellow one is natural protein sequences and the green one is generated protein sequences.  b. Shannon's entropy at each position of the multiple sequence alignment results. c. Multiple sequence alignment results of 5 generated lysozyme sequences and 5 natural lysozyme sequences randomly sampled from both lysozyme C and lysozyme G.}
    \label{result}
\end{figure*}

Additionally, MMseqs2 is utilized for clustering analysis at the three-dimensional structural level. The comprehensive results of our analysis of the generated lysozyme sequences are presented and discussed in Figure \ref{result}.

Generative models aim to capture the conditional distribution of real data. Figure \ref{result}a illustrates the two-dimensional t-SNE visualization derived from a validation dataset comprising approximately 500 natural lysozyme sequences, with an equivalent number of generated lysozyme sequences. The x-axis represents the first dimension of t-SNE, while the y-axis corresponds to the second dimension. In this visualization, the real distribution of natural lysozyme sequences is depicted in yellow, and the generated distribution of lysozyme sequences is shown in green.

For the t-SNE analysis, we enhance the representation of both real and generated data by employing the ESM-2 model to reduce the dimensionality of the data. We set the number of components, n\_components, to 2, and retain the default settings for other parameters, which are unlikely to significantly affect the t-SNE results. Our findings indicate that the distribution of the generated lysozyme sequences closely resembles that of the real sequences, suggesting that the generative model has successfully learned the underlying distribution of the data.

 Figure \ref{result}b presents the Shannon entropy calculated for each position in a multiple sequence alignment (MSA) that includes 6 natural lysozyme sequences and 6 synthetically generated lysozyme sequences. To conduct this analysis, we utilized Clustal Omega to perform the MSA, incorporating both lysozyme C and lysozyme G for validation purposes. The outcomes of the MSA, depicted in Figure \ref{result}c, suggest that the sequences are homologous. Further analysis involved calculating Shannon's entropy for the MSA results at each position, which reflects the frequency or probabilities within the alignment. The entropy plot in Figure \ref{result}b reveals that the entropy levels of both natural and generated sequences are comparable, indicating that our model is capable of learning the conditional distribution and retaining the characteristic features of lysozyme sequences.

\begin{figure}
    \centering
    \includegraphics[width=\textwidth]{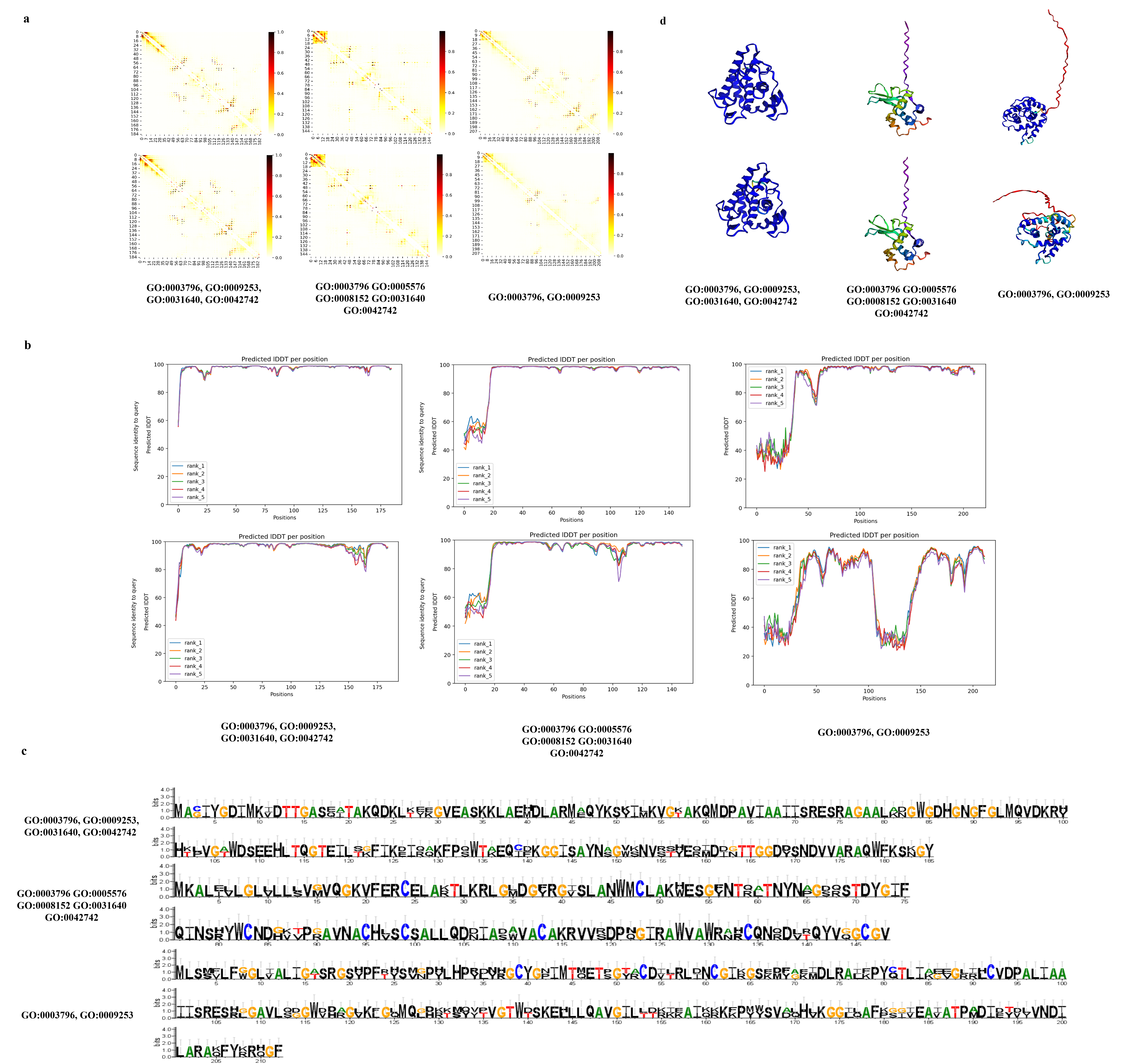}
    \caption{ a. The attention map for the natural lysozyme above
     and the generated lysozyme below.
A larger value indicates that the amino acid residue at a position on the x-axis is in contact with the amino acid residue at a position on the y-axis.
b. The IDDT value
for each position: the natural lysozyme above and the generated lysozyme below.
c. The results of the conservation analysis based on MSA for one natural lysozyme and one generated lysozyme, highlighting different gene ontology terms.
d. Detailed 3D structure predictions by AlphaFold2, with the natural structure above and the generated prediction below.
}
    \label{4go}
\end{figure}

 The detailed predictions for the 2D and 3D structures of these sequences are analyzed in Figure \ref{4go}.
 In Figure \ref{4go} (subfigures a, b, c, d), the results for the real lysozyme sequences are displayed above, while those for the generated sequences are below. Figure \ref{4go} presents the analysis of lysozyme functionality with different Gene Ontology (GO) terms. We employed ESM-2 for predicting the two-dimensional structures. The attention maps of two lysozyme sequences shown in Figures \ref{4go} illustrate the protein sequences' structure through heatmaps, where darker colors indicate higher values, signifying contact between amino acid residues at specific positions. This contact information aids in inferring the 2D and 3D structures of the proteins.  Figures \ref{4go}b displays the IDDT (Identification of Directly Determined Contacts \citep{Weigt2009} results for each position, comparing the natural and generated sequences. The significance of these metrics is elaborated in Section 2.5. From these results, it is evident that the generated lysozyme sequences retain most of the structural information.

 Particularly, Figures \ref{4go}c presents the results of conservation analysis. Here, the x-axis represents the amino acids from the MSA results of one natural and one generated lysozyme sequence. The y-axis indicates the probability of occurrence of each amino acid, with the height of each bar being proportional to the frequency of occurrence of the corresponding residue at that position, measured in bits. This analysis provides insights into the conservation of amino acid residues within the sequences. Lastly, Detailed 3D structure predictions by AlphaFold2 are shown in Figure \ref{4go}d, where a visual comparison reveals a high degree of similarity between the generated and natural protein structures. Additional results are given
 in the Supplementary Materials.

\subsection{Malate dehydrogenase (MDH) sequences}
In our second example, we focus on Malate dehydrogenase (MDH) protein sequences. MDH is a crucial enzyme in cellular metabolism that catalyzes the reversible oxidation of malate to oxaloacetate. In ProteinRG, we include 11 GO annotations related to MDH enzymes as input labels. The process for training the ProteinGR model with MDH sequences follows the same procedure as described for the Lysozyme sequences above.

\begin{figure*}
	\centering
	\includegraphics[width=\textwidth]{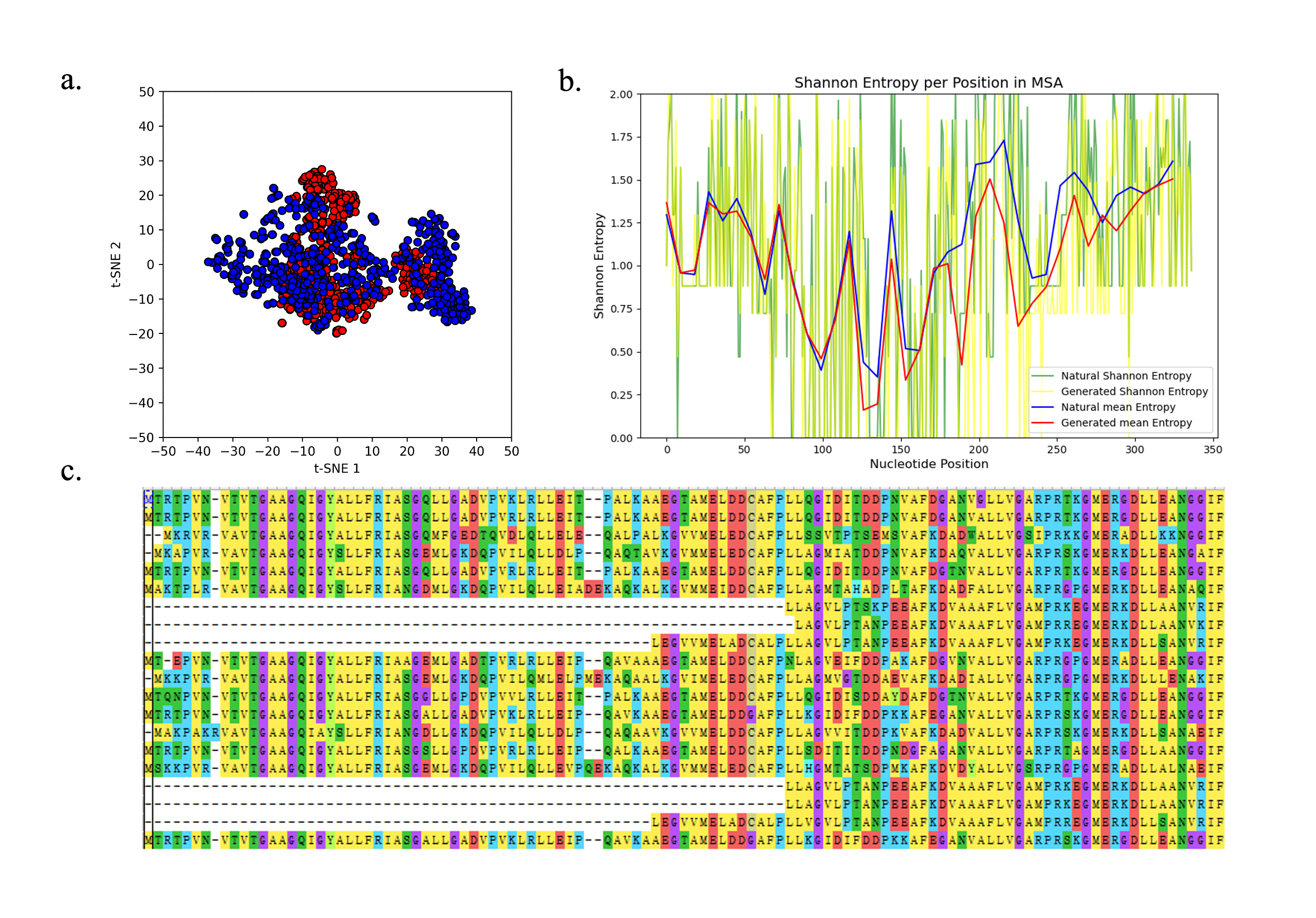}
	\caption{The general analysis results of the natural MDH sequence and the generated MDH sequence. a. The t-SNE result about 500 natural MDH sequences and 500 generated MDH sequences. b. The shannon entropy result with 10 natural MDH sequences and 10 generated MDH sequences. c. The multiple sequence alignment result by MEGA software.}
	\label{mdh_result}
\end{figure*}

\begin{figure}[H]
    \centering
    \includegraphics[width=\textwidth]{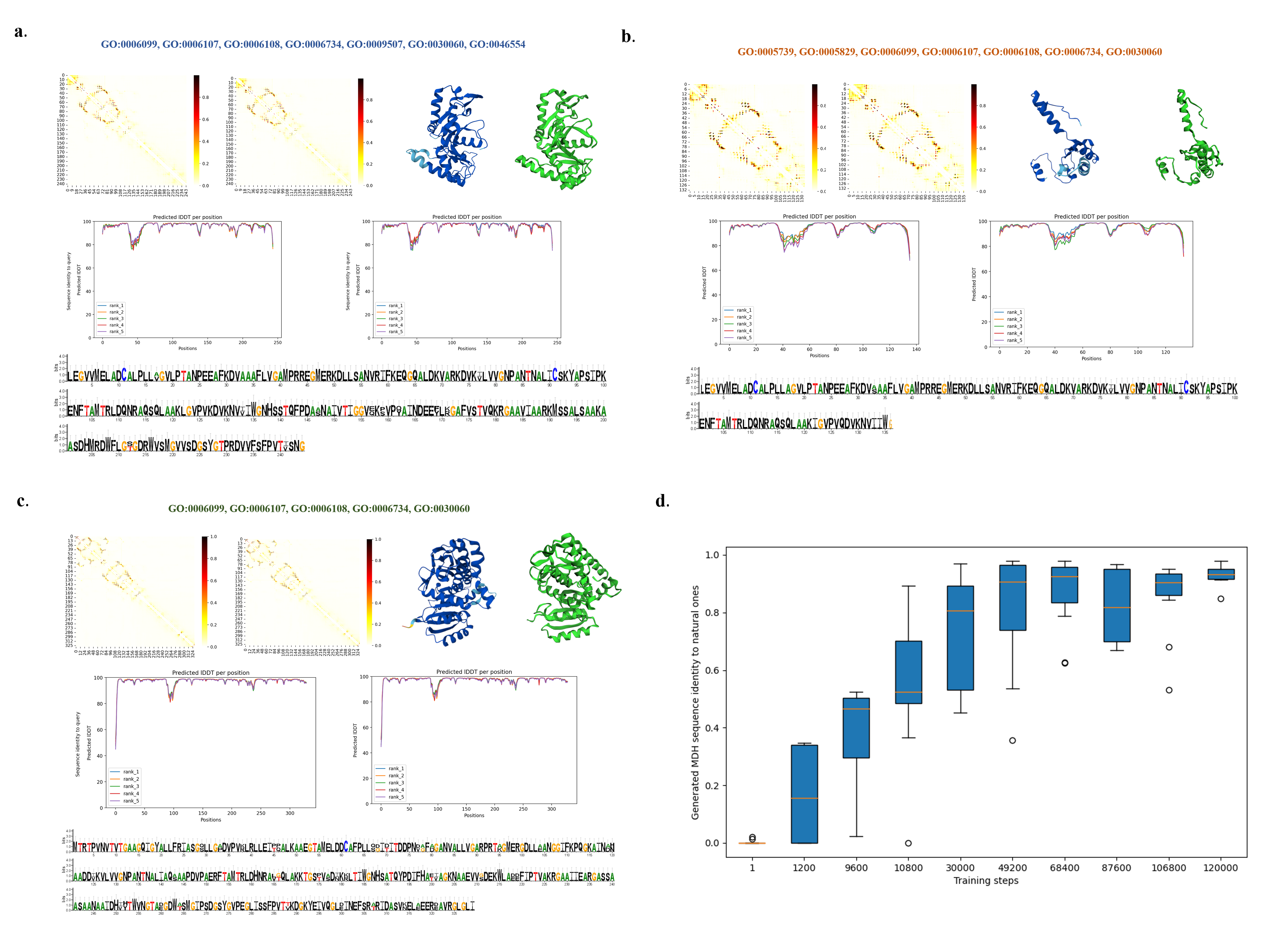}
    \caption{The analysis results of the natural MDH sequence (left) and the generated MDH sequence (right). The attention map, IDDT at each position, conversation analysis and 3D-structure prediction from Alphafold2 are given in \textbf{a}, \textbf{b}, and \textbf{c}.
     \textbf{a}. The input GO annotations for the protein sequence include GO:0006099, 0006107, 0006108, 0006734, 0009507, 0030060, and 0046554. \textbf{b}. The input GO annotations include GO:0005739, 0005829, 0006099,  0006107, 0006108, 0006734, and 0030060. \textbf{c}. The input GO annotations include GO:0006099, 0006107, 0006108, 0006734, and 0030060.
     \textbf{d}. The similarity of the 32 generated MDH sequences to the natural ones from test data at different training steps.}
    \label{mdh}
\end{figure}

Figures \ref{mdh_result} show the general analysis result about the natural MDH sequences and generated MHD sequences, which contains t-SNE result in figure \ref{mdh_result}a, the shannon entropy result with 20 sequences in figure \ref{mdh_result}b, and multiple sequence alignment result in figure \ref{mdh_result}c. Figures \ref{mdh}a,b,c present the analysis results of the natural MDH sequence on the left and the generated MDH sequence on the right. The attention map that shows the 2D interactions between each amino acid of the sequence is predicted by ESM-2. The IDDT scores and 3D structure prediction are conducted by Alphafold2. The sequence conversation analysis is based on the MSA and MEGA for one natural MDH sequence and one generated MDH sequence with the same GO annotations.

Figures \ref{mdh}a, \ref{mdh}b, and \ref{mdh}c present the analysis results, with the natural MDH sequence on the left and the generated MDH sequence on the right. The attention map, showing the 2D interactions between each amino acid of the sequence, is predicted by ESM-2. The IDDT scores and 3D structure predictions are conducted by AlphaFold2. Sequence conservation analysis is based on MSA and MEGA for one natural MDH sequence and one generated MDH sequence with the same GO annotations.

For Figure \ref{mdh}a, the input GO annotations include:
Tricarboxylic acid cycle (GO:0006099),
Oxaloacetate metabolic process (GO:0006107),
Malate metabolic process (GO:0006108),
NADH metabolic process (GO:0006734),
Chloroplast (GO:0009507),
L-malate dehydrogenase activity (GO:0030060),
Malate dehydrogenase activity (GO:0046554).
For Figure \ref{mdh}b, the input GO annotations include:
Mitochondrion (GO:0005739),
Cytosol (GO:0005829),
Tricarboxylic acid cycle (GO:0006099),
Oxaloacetate metabolic process (GO:0006107),
Malate metabolic process (GO:0006108),
NADH metabolic process (GO:0006734).
For Figure \ref{mdh}c, the input GO annotations include:
Tricarboxylic acid cycle (GO:0006099),
Oxaloacetate metabolic process (GO:0006107),
Malate metabolic process (GO:0006108),
NADH metabolic process (GO:0006734).

Additionally, Figure \ref{mdh}d presents the sequence identity of the generated MDH sequences to the natural MDH sequences from the test dataset at different training steps. The number of selected generated sequences is 32. We calculate the MDH sequence identity using the Biopython package (\cite{cock2009biopython}). The final identity of the generated MDH sequences can reach up to 95\%.

\subsection{Comparative analysis: diverse protein sequences}
We conducted a comparative analysis of our model against both large language models for sequence data and deep-learning models for protein sequence generation.

 We compare our method with the following methods.
\begin{itemize}
    \item CVAE \citep{greener2018design}: This model is a Conditional Variational Autoencoder designed for protein sequences. We adapted the model to accommodate the 50 labels pertinent to our problem setting and conducted a Bayesian optimization search for the best hyperparameters.
    \item ProGen \citep{madani2023large}:  ProGen is a language model leveraging a state-of-the-art Transformer architecture. Conditional information is integrated by prefixing label tokens to the sequence. We scaled down the model size and retrained it using our dataset.
    \item One-per-label GAN (OpL-GAN): OpL-GAN is an unconditional protein generation method. Therefore, in using this method, we consider one instance of ProteoGAN for each label, omitting the conditioning mechanism (resulting in a total of 50 models). Sequences for a specific label are generated by sampling from the GAN trained exclusively on sequences annotated with that label. This model allows us to evaluate whether training 50 distinct models could be an alternative to employing a conditioning mechanism.
    \item ProteoGAN \citep{kucera2022conditional}: A conditional GAN tailored for the functional design of proteins, utilizing hierarchical GO annotations. Unlike other models, ProteoGAN takes discrete labels as input rather than representations.
\end{itemize}

Since these methods are designed for protein sequence data with large sample sizes, they are not suitable for the Lysozyme and MDH datasets used in our analysis, as the sample sizes of these two datasets are too small for these existing methods to be effective. Therefore, we use the protein sequence dataset included in \cite{kucera2022conditional}, which consists of 157,890 protein sequences, each annotated with one of 50 distinct GO functions.

The performance metrics, including Maximum Mean Discrepancy (MMD), Mean Reciprocal Rank (MRR), Entropy, and Distance, are detailed in Table \ref{table}. The training process is shown in Figure S2 in the Supplementary Materials. For Entropy and Distance, we report the average entropy across feature dimensions and the average pairwise Reproducing Kernel Hilbert Space (RKHS) distance between sequences, respectively. Based on these metrics, our
analysis indicates that our model outperforms these existing models in generating protein sequences.

\begin{table}[H]
    \centering
    \setlength{\tabcolsep}{8mm}{
    \begin{tabular}{@{}ccccc@{}}
    \toprule
    Model  &  MMD  &  MRR    &  Entropy  &  Distance \\ \midrule
    OpL\-GAN  & 0.036 &  0.597 & \-0.062 & 0.022 \\
    ProGen  & 0.048  & 0.394 & \-0.156 & 0.037 \\
    CVAE  & 0.232  & 0.301  &0.247 & 0.145 \\
    ProteoGAN   & 0.043  & 0.554   & \-0.010 & 0.012\\
    Our model  & \textbf{0.034}  & \textbf{0.603}  &\textbf{0.006} & \textbf{0.001}
    \\ \bottomrule
    \end{tabular}}
\caption{Comparison the performance of our model with the existing methods OpL\-GAN, ProGen, CVAE, and ProteoGAN in terms of MMD, MRR, Entropy, and Distance.}
\label{table}
\end{table}

We have selected five distinct types of protein sequences from the dataset, each associated with different functions, to demonstrate the learning capabilities of our model with limited data. Each type of protein sequence is represented by approximately one thousand samples for training purposes.

\begin{figure*}
    \centering
    \includegraphics[width=1\linewidth]{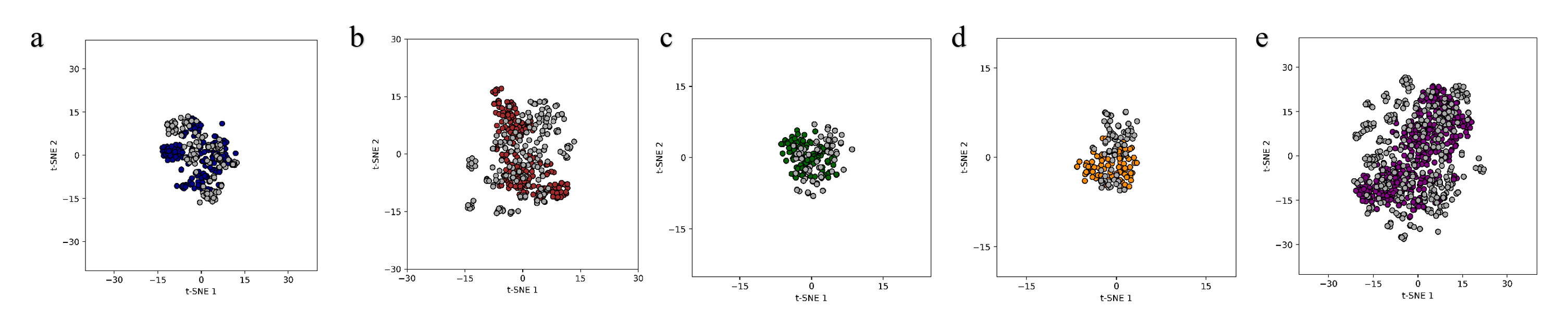}
    \caption{t-SNE visualization from natural protein and generated protein with five different functions. a. Drug binding and transmembrane transporter activity. b. Catalytic activity, cofactor binding, and anion binding. c. Small molecule binding, signaling receptor activity, catalytic activity, acting on a protein and kinase activity. d. Ion transmembrane transporter activity and cation binding. e. DNA binding and RNA binding.}
    \label{5kind}
\end{figure*}

Figure \ref{5kind}a illustrates proteins involved in drug binding and transmembrane transporter activity. Figure \ref{5kind}b showcases proteins with catalytic activity, cofactor binding, and anion binding functions. Figure \ref{5kind}c highlights proteins that are associated with small molecule binding, signaling receptor activity, catalytic activity with a focus on proteins and kinase activity. Figure \ref{5kind}d presents proteins functioning in ion transmembrane transporter activity and cation binding. Lastly, Figure \ref{5kind}e depicts proteins with DNA binding and RNA binding capabilities. Figure \ref{5kind}f provides an overview of the distribution of these five types of protein sequences, offering insights into the diversity and specialization of functions within the dataset.

\subsection{Evaluation of the analysis procedure}
\begin{figure}
    \centering
    \includegraphics[width=1\linewidth]{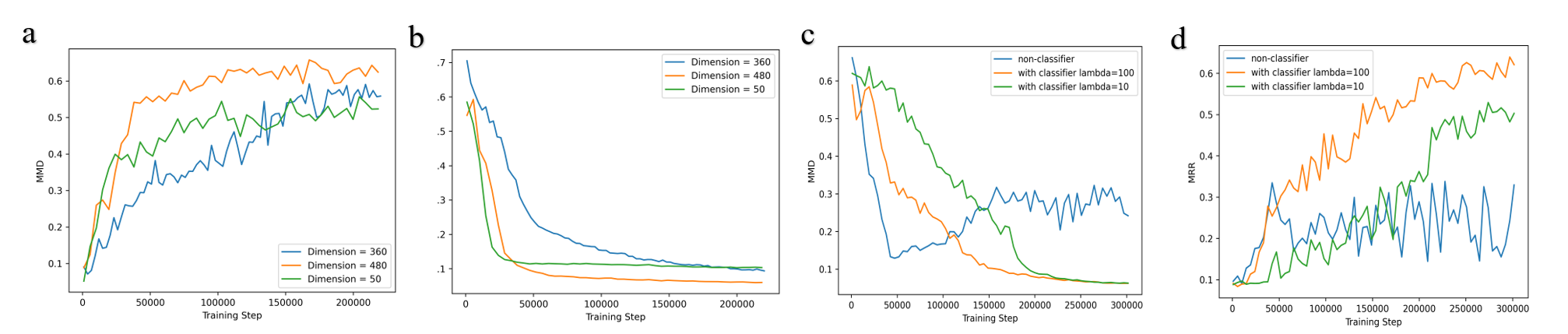}
    \caption{The influence of dimension of latent representation $r$
    and hyperparameter $\beta$ on MRR and MMD.   a. The MRR with different $r$ values versus training steps;  b. The MMD with different $r$ values versus training steps;
    c. The MRR with different $\beta$ values versus training steps; d. The MMD with different $\beta$ values versus training steps. }
    \label{dim}
\end{figure}

We first examine the influence of varying dimensions of latent representation on the outcomes of our model. For a detailed account of the results, please refer to Figure \ref{dim}. In Figure \ref{dim}a, we present the Mean Reciprocal Rank (MRR) values as a function of the training steps. Similarly, Figure \ref{dim}b illustrates the Maximum Mean Discrepancy (MMD) values over the course of training. From these figures, we observe that the model's performance improves with an increase in the dimensionality of the latent representation. Additionally, when comparing performance based on label
conditions instead of a representation with a dimension of 50, we observe a faster convergence rate but reduced quality.

However, due to constraints in computational resources, we were unable to test the model with significantly higher dimensions of latent representation. We hypothesize that there may exist a threshold dimension beyond which the performance gains may plateau or diminish. Further investigation into this aspect could provide valuable insights into the optimal dimensionality for latent representations in our model.

In our analysis, we also explore the necessity of incorporating a discriminative classifier into our model. To do this, we experiment with three different orders of magnitude for the hyperparameter $\beta$. Our findings indicate that increasing $\beta$ tenfold enhances the performance of our model, suggesting that the discriminative classifier plays a crucial role in improving model efficacy. However, we observe a decline in performance when $\beta$ exceeds 300, identifying this as a threshold beyond which the benefits of increasing $\beta$ diminish.

To optimize the value of $\beta$, we employ an optimization package \citep{falkner2018bohb}, ultimately determining that a $\beta$ value of 175 yields the best performance. This optimization process and the comparative analysis of different $\beta$ values are visually represented in Figure \ref{dim} c and d. In this figure, the x-axis represents the training steps, while the y-axis denotes the metrics, including Maximum Mean Discrepancy (MMD) and Mean Reciprocal Rank (MRR), as previously mentioned. Specifically, Figure\ref{dim}a illustrates the variation of the MRR value across training steps, and Figure\ref{dim}b displays the changes in the MMD value over the course of training.

In our study, we also investigate the potential benefits of fine-tuning ESM-2. Based on our metrics, we observe that fine-tuning ESM-2 does not significantly affect the quality of the generated sequences. This could be attributed to the fact that ESM-2 has been pre-trained on millions of protein sequences, and it is likely that our training dataset might overlap with the pre-trained dataset. Consequently, the representations derived from the pre-trained model might already encapsulate functional information about the protein sequences, potentially offering a broader perspective than that provided by Gene Ontology (GO) annotations alone.

However, our investigation does not delve into optimizing the fine-tuning process for the pre-trained model. There exists a possibility that enhancing the fine-tuning methodology for ESM-2 could yield different outcomes. Therefore, future research might benefit from exploring more sophisticated approaches to fine-tuning ESM-2, which could potentially lead to improved performance and more insightful conclusions.

\section{Conclusion}
\label{sec:conc}
In this work, we have proposed ProteinRG, a novel hierarchical model that synergistically integrates a pre-trained large language model with a generative model to facilitate the design of functional proteins when data is scarce. Our model is structured into three distinct modules: (a) the latent feature capture module, (b) the first-stage generative module, and (c) the second-stage generative module. Initially, the model generates a latent representation of a protein sequence, which is subsequently used to produce a protein sequence endowed with specific functions.

We have applied ProteinRG to a variety of functional protein sequences and evaluated the generated outcomes from three different perspectives: multiple sequence alignment, t-SNE distribution analysis, and 3-D structures. The findings indicate that, even with a limited amount of data,  ProteinRG can simultaneously ensure similarity and condition consistency in the generated protein sequences, demonstrating superior performance in comparison to other generative models for protein sequence design. Additionally, the fundamental concept of employing a large model for learning data representation, dimension reduction,  and hierarchical generative modeling is broadly applicable and can be extended to other challenges, such as modeling RNA data or any type of sequence data.

Despite these promising results, our approach has certain limitations. One limitation arises when generating diverse protein families concurrently; the Gene Ontology (GO) annotations database must contain a substantial number of labels to facilitate effective embedding and classification. This requirement can pose challenges related to data completeness and may incur additional costs due to the increased dimensionality of the input and output for classification. Additionally, there may be a need to refine the method of fine-tuning the ESM-2 model during the latent feature extraction stage to further enhance performance. In specific applications, such as studies targeting certain drugs through protein sequence generation, laboratory experiments are necessary to validate the generated results. While this work focuses on protein sequence data, our proposed generative model is also applicable to other sequence data, such as mRNAs. We plan to explore these aspects in future work.

{\singlespace
\bibliographystyle{apalike} 
\bibliography{prg_bib}
}
\end{document}